\documentclass[aps,prl,superscriptaddress,preprintnumbers,twocolumn,groupedaddress,showpacs]{revtex4}

\usepackage{graphicx}
\usepackage{hyperref}
\usepackage{amsmath}

\newcommand{\rT}{{\mathrm{T}}}

\newcommand{\LO}{{\mathrm{LO}}}
\newcommand{\NLO}{{\mathrm{NLO}}}
\newcommand{\EW}{{\mathrm{EW}}}

\def\mathswitchr#1{\relax\ifmmode{\mathrm{#1}}\else$\mathrm{#1}$\fi}
\newcommand{\PZ}{\mathswitchr Z}
\newcommand{\PH}{\mathswitchr H}

\newcommand{\Pe}{\mathswitchr e}

\newcommand{\Pp}{\mathswitchr p}
\newcommand{\Pj}{\mathswitchr j}

\newcommand{\Pt}{\mathswitchr t}
\newcommand{\Pu}{\mathswitchr u}
\newcommand{\Pd}{\mathswitchr d}
\newcommand{\PW}{\mathswitchr W}
\newcommand{\Pq}{\mathswitch{q}}
\newcommand{\Pl}{\ell}

\def\mathswitch#1{\relax\ifmmode#1\else$#1$\fi}

\newcommand{\MW}{\mathswitch {M_\PW}}

\newcommand{\MZ}{\mathswitch {M_\PZ}}
\newcommand{\MH}{\mathswitch {M_\PH}}

\newcommand{\Mt}{\mathswitch {m_\Pt}}
\newcommand{\GW}{\mathswitch {\Gamma_\PW}}
\newcommand{\GZ}{\mathswitch {\Gamma_\PZ}}
\newcommand{\GH}{\mathswitch {\Gamma_\PH}}
\newcommand{\Gt}{\mathswitch {\Gamma_\Pt}}

\newcommand{\TeV}{\unskip\,\mathrm{TeV}}
\newcommand{\GeV}{\unskip\,\mathrm{GeV}}

\newcommand{\fb}{\unskip\,\mathrm{fb}}

\def\refeq#1{\mbox{(\ref{#1})}}

\def\reffi#1{\mbox{Fig.~\ref{#1}}}

\def\refta#1{\mbox{Table~\ref{#1}}}

\def\citere#1{\mbox{Ref.~\cite{#1}}}
\def\citeres#1{\mbox{Refs.~\cite{#1}}}

\newcommand{\ie}{\textit{i.e.}}

\def\draftdate{\relax}
\def\mda{\relax}
\def\mua{\relax}
\def\mla{\relax}
\def\draft{
\def\thtystars{******************************}
\def\sixtystars{\thtystars\thtystars}
\typeout{}
\typeout{\sixtystars**}
\typeout{* Draft mode!
         For final version remove \protect\draft\space in source file *}
\typeout{\sixtystars**}
\typeout{}
\def\draftdate{\today}
\def\mua{\marginpar[\boldmath\hfill$\uparrow$]%
                   {\boldmath$\uparrow$\hfill}%
                    \typeout{marginpar: $\uparrow$}\ignorespaces}
\def\mda{\marginpar[\boldmath\hfill$\downarrow$]%
                   {\boldmath$\downarrow$\hfill}%
                    \typeout{marginpar: $\downarrow$}\ignorespaces}
\def\mla{\marginpar[\boldmath\hfill$\rightarrow$]%
                   {\boldmath$\leftarrow$\hfill}%
                    \typeout{marginpar: $\leftrightarrow$}\ignorespaces}
\def\Mua{\marginpar[\boldmath\hfill$\Uparrow$]%
                   {\boldmath$\Uparrow$\hfill}%
                    \typeout{marginpar: $\uparrow$}\ignorespaces}
\def\Mda{\marginpar[\boldmath\hfill$\Downarrow$]%
                   {\boldmath$\Downarrow$\hfill}%
                    \typeout{marginpar: $\downarrow$}\ignorespaces}
\def\Mla{\marginpar[\boldmath\hfill$\Rightarrow$]%
                   {\boldmath$\Leftarrow $\hfill}%
                    \typeout{marginpar: $\leftrightarrow$}\ignorespaces}
\overfullrule 5pt
\oddsidemargin -10mm
\marginparwidth 15mm
}

\draft

\begin{document}

\title{\boldmath{Large electroweak corrections to vector-boson
    scattering at the Large Hadron Collider}}
 
\author{Benedikt~Biedermann}
\affiliation{Universit\"at W\"urzburg,
Institut f\"ur Theoretische Physik und Astrophysik, 
D-97074 W\"urzburg, Germany}

\author{Ansgar~Denner}
\affiliation{Universit\"at W\"urzburg,
Institut f\"ur Theoretische Physik und Astrophysik, 
D-97074 W\"urzburg, Germany}

\author{Mathieu~Pellen}
\affiliation{Universit\"at W\"urzburg,
Institut f\"ur Theoretische Physik und Astrophysik, 
D-97074 W\"urzburg, Germany}

\date{\today}

\begin{abstract}
  For the first time full next-to-leading-order electroweak
  corrections to off-shell vector-boson scattering are presented.  The
  computation features the complete matrix elements, including all
  non-resonant and off-shell contributions, to the
  electroweak process $\Pp \Pp \to \mu^+ \nu_\mu \Pe^+ \nu_{\Pe} \Pj
  \Pj$ and is fully differential. We find  surprisingly
  large corrections, reaching $-16\%$ for the fiducial cross section,
  as an intrinsic feature of vector-boson-scattering processes. We elucidate
the origin of these large electroweak corrections upon using the
double-pole approximation and the effective vector-boson approximation
along with leading-logarithmic corrections.
\end{abstract}


\maketitle

\subsection{Introduction}

Since the Higgs boson has been discovered at the Large Hadron Collider
(LHC), a significant experimental program has been devoted to the
study of the electroweak (EW) sector of the Standard Model
\cite{Green:2016trm}. The Higgs boson plays a crucial role in
vector-boson scattering (VBS) as it prevents the amplitude from
violating unitarity at high energies.  VBS is thus a basic process to
investigate the mechanism of EW symmetry breaking.  In
addition, VBS is a key testing ground for possible new interactions,
as it is particularly sensitive to small deviations from the Standard
Model, which can, for example, be parametrized by anomalous quartic
gauge-boson couplings. Hence, precise predictions for this process
will allow one to impose more stringent exclusion limits on new physics
models or even trigger the discovery of new fundamental mechanisms.

VBS at the LHC is an exclusive process where a constituent of each
colliding proton emits a weak vector boson $V=\PW,\PZ$ which then
scatter off each other. The emissions from the protons cause jets in
the forward and backward directions with large rapidity difference and
dijet invariant mass. The resulting $VV\Pj\Pj$ final state receives
contributions from both EW and QCD mediated processes, referred to as
EW and strong production that can be separated in a gauge-invariant
way.  Strong production can be suppressed by requiring the
above-mentioned tagging jets in the forward and backward directions.
In the following, we refer to the EW production mode of the $VV\Pj\Pj$
final state as the actual VBS process.

The present letter focuses on the EW production of two off-shell ${\rm
  W^+}$ bosons in association with two jets, \ie\ $\Pp \Pp \to \mu^+
\nu_\mu \Pe^+ \nu_{\Pe} \Pj \Pj$, which has been identified as the
most promising channel for the measurement of VBS
\cite{Campbell:2015vwa} at the LHC.  For like-sign $\PW\PW$
scattering, the strong production mode does not dominate over the EW
mode, in contrast to most other VBS processes.  Evidence
for this process has already been reported by both the ATLAS
\cite{Aad:2014zda,Aaboud:2016ffv} and CMS \cite{Khachatryan:2014sta}
collaborations.

Significant interest has been devoted in the past to the computation
of higher-order corrections to VBS and its main irreducible background
processes
\cite{Melia:2010bm,Melia:2011dw,Jager:2009xx,Jager:2011ms,Denner:2012dz,Campanario:2013gea,Baglio:2014uba,Rauch:2016pai}.
So far, these computations have focused exclusively on next-to-leading-order (NLO) QCD
corrections, and no NLO EW computation has been performed yet.  As the
impact of EW corrections grows with the energy owing to the presence
of logarithms of the ratio of energy and EW gauge-boson mass,
so-called Sudakov logarithms, they have to be calculated in view of
the increasing energy and luminosity of the LHC. In this letter we
present first results for the  full NLO EW
corrections to the off-shell VBS process $\Pp \Pp \to \mu^+ \nu_\mu
\Pe^+ \nu_{\Pe} \Pj \Pj$.

The EW corrections to the integrated cross section turn out to be
surprisingly large compared to those for other standard LHC processes.
In order to elucidate the origin of these enhanced corrections, we
study results for EW corrections also in two approximations, the
double-pole approximation and the effective vector-boson
approximation (EVBA).  Our results allow one to include these
corrections in precise measurements of VBS at the LHC.

\subsection{Details of the calculation}

We consider the leading-order EW process $\Pp \Pp \to \mu^+ \nu_\mu
\Pe^+ \nu_{\Pe} \Pj \Pj$ of order $\mathcal{O}{\left(
    \alpha^{6}\right)}$.  The dominant partonic channel $\Pu \Pu \to
\mu^+ \nu_\mu \Pe^+ \nu_{\Pe} \Pd \Pd$ accounts for about two thirds
of the cross section.  The second largest channel $\Pu \overline{\Pd}
\to \mu^+ \nu_\mu \Pe^+ \nu_{\Pe} \Pd \overline{\Pu}$ features a Higgs
boson in the $s$~channel and makes up $16\%$ of the cross section.
The remaining partonic channels sum up to $17\%$ of the cross section.
The EW NLO corrections comprise all contributions of order
$\mathcal{O}{\left(\alpha^{7}\right)}$.  These include the complete EW
virtual one-loop amplitude and real photon radiation, \ie\ the process
$\Pp \Pp \to \mu^+ \nu_\mu \Pe^+ \nu_{\Pe} \Pj \Pj\gamma$.  At order
$\mathcal{O}{\left(\alpha^{7}\right)}$, also contributions of the type
$\Pq_1 \gamma \to \mu^+ \nu_\mu \Pe^+ \nu_{\Pe} \Pq_2 \Pq_3
\bar{\Pq}_4$ appear, where $\Pq_i$ are quarks of possibly different
type.  These real corrections are suppressed owing to the photon
distribution function and therefore have been omitted in the present
computation.

The resonant particles are treated within the complex-mass
scheme~\cite{Denner:1999gp,Denner:2005fg}, ensuring gauge invariance.  To evaluate all one-loop
amplitudes in the 6-body phase space, the computer code {\sc
  Recola}~\cite{Actis:2012qn,Actis:2016mpe} and the {\sc Collier}
library \cite{Denner:2014gla,Denner:2016kdg} are employed.  The
phase-space integration is carried out with two different Monte Carlo
programs that have been used in
\citeres{Biedermann:2016yvs,Biedermann:2016guo} and
\citeres{Denner:2015yca,Denner:2016jyo}, respectively.  The infrared
singularities are treated via the Catani--Seymour dipole subtraction
formalism \cite{Catani:1996vz,Dittmaier:1999mb}.  The collinear
initial-state splittings are handled within the DIS factorization
scheme following 
\citeres{Diener:2005me,Dittmaier:2009cr}.  

To ensure the correctness of the results, a number of checks has been
performed.  We have verified numerically that the sum of all
corrections is infrared finite.  The hadronic Born cross section has
been compared against the computer code {\sc\small
  MadGraph5\_aMC@NLO}~\cite{Alwall:2014hca}, which has also been used
to check the tree-level matrix elements squared (for Born and real
radiation).  Finally, for the dominant partonic channels ($\Pu\Pu$ and
$\Pu\bar\Pd$) a computation in the double-pole approximation (based on
an automatized implementation following the one of
\citere{Denner:2016jyo}) has confirmed the NLO EW corrections obtained
in the full calculation within an accuracy below $1\%$.

\subsection{Input parameters and event selections}

We present theoretical predictions for the LHC at the center-of-mass
energy of $13\TeV$.  The on-shell values for the masses and widths of
the gauge bosons 
\begin{equation}
\begin{array}[b]{rcl@{\quad}rcl}
  \MW^{\rm os} &=& 80.385  \GeV, & \GW^{\rm os} &=& 2.085 \GeV, \\
  \MZ^{\rm os} &=& 91.1876 \GeV, & \GZ^{\rm os} &=& 2.4952\GeV
\end{array}
\end{equation}
are converted into pole masses according to
\begin{eqnarray}
&& M_V = M_V^{\rm os}/c_V, \quad \Gamma_V = \Gamma_V^{\rm os}/c_V, \nonumber\\
&& c_V=\sqrt{1+(\Gamma_V^{\rm os}/M_V^{\rm os})^2}, \quad V=\PW,\PZ.
\end{eqnarray}
The Higgs-boson and top-quark masses and widths are fixed to 
\begin{equation}
\begin{array}[b]{rcl@{\quad}rcl}
\MH &=& 125\GeV,         & \GH   &=& 4.07 \times 10^{-3}\GeV, \nonumber \\
\Mt &=& 173.21\GeV,         & \Gt   &=& 0\GeV .
\end{array}
\end{equation}
The top-quark width can be neglected since no resonant top quarks
appear in the matrix elements.

For the electromagnetic coupling $\alpha$, the $G_\mu$ scheme is
used where $\alpha$ is obtained from the Fermi constant,
\begin{equation}
\alpha_{G_\mu} = \sqrt{2}G_\mu\MW^2\left(1-\MW^2/\MZ^2\right)/\pi ,
\end{equation}
with
\begin{equation}
G_\mu= 1.16637\times 10^{-5} \GeV^{-2}.
\end{equation}

We have chosen the set of parton distribution functions
NNPDF3.0QED~\cite{Ball:2013hta,Ball:2014uwa}.  The renormalization and
factorization scales, $\mu_{\mathrm{ren}}$ and $\mu_{\mathrm{fact}}$,
are set equal to the pole mass of the W~boson,
$\mu_{\mathrm{ren}}=\mu_{\mathrm{fact}}=\MW$.

The event selection for VBS is based on the experimental analyses of
ATLAS \cite{Aad:2014zda} and CMS \cite{Khachatryan:2014sta}.  Quarks
and gluons are clustered using the anti-$k_\rT$ algorithm
\cite{Cacciari:2008gp} with jet-resolution parameter $R=0.4$.  The
recombination of photons with charged particles employs a resolution
parameter $R=0.1$.

For each jet and charged lepton, a cut on its transverse momentum and
its rapidity is applied,
\begin{eqnarray}
\label{eq:jetcut}
 p_{\rT,\rm j} &{}>& 30\GeV, \quad |y_{\rm j}| < 4.5,\\
\label{eq:leptoncut}
 p_{\rT,\ell} &{}>& 20\GeV, \quad |y_{\ell}| < 2.5,
\end{eqnarray}
and the missing energy has to fulfill
\begin{equation}\label{eq:emisscut}
 E^{\mathrm{miss}}_{\rT} > 40\GeV.
\end{equation}
For the pair of jets, an invariant mass cut and a cut on the difference of the
rapidities is applied,
\begin{equation}\label{eq:jjcuts}
 M_{\Pj \Pj} > 500\GeV, \quad |\Delta y_{\Pj \Pj}| > 2.5.
\end{equation}
Finally, the leptons are required to be isolated,
\begin{equation}\label{eq:drcut}
 \Delta R_{\Pl\Pl} > 0.3, \quad \Delta R_{\Pj\Pl} > 0.3,
\end{equation}
where
\begin{equation}
\Delta R_{ij} = \sqrt{(\Delta y_{ij})^2+(\Delta \phi_{ij})^2}
\end{equation}
is the rapidity--azimuthal-angle distance of the objects $i$ and $j$.

\subsection{Numerical results}

The fiducial cross section for the VBS event
selection \refeq{eq:jetcut}--\refeq{eq:drcut}
and the corresponding NLO EW corrections are reported in \refta{tab:xsec}.
\begin{table}
\begin{center}
\begin{tabular}
{|ccc|}
\hline
$\sigma^{\LO}$~[fb] &  $\sigma^{\NLO}_{\EW}$~[fb] & $\delta_{\EW}~[\%]$ 
\\
\hline
$\phantom{1}1.5348(2)$ & $\phantom{1}1.2895(6)$& $-16.0$ 
\\
\hline
\end{tabular}
\end{center}
\caption{LO and NLO cross section for $\Pp \Pp \to \mu^+ \nu_\mu \Pe^+ \nu_{\Pe} \Pj\Pj$ at $13\TeV$ at the LHC.
The corresponding EW corrections are given in per cent.
The digit in parenthesis indicates the integration error.}
\label{tab:xsec}
\end{table}
Strikingly, the EW corrections turn out to be $-16\%$ and thus
surprisingly large for a fiducial cross section.  For typical LHC
processes, such large negative corrections originating from Sudakov
logarithms generically show up in the high-energy tails of
distributions which usually do not dominate the integrated cross
section.

Moreover, these large corrections are not due to the VBS event selections
described above but are intrinsic to the VBS process at the
LHC \footnote{Even for a center-of-mass energy of $\sqrt{s}=7\TeV$ we
  find EW corrections of $-13\%$ for the dominating channels.}.
Indeed, dropping the cuts \refeq{eq:jjcuts} on the two jets and on the
missing energy \refeq{eq:emisscut} and relaxing the requirements on
the transverse momenta \refeq{eq:jetcut} and \refeq{eq:leptoncut}
leave the corrections at the same level.

\begin{figure}
\hspace{-2cm}
\includegraphics[width=.47\textwidth]{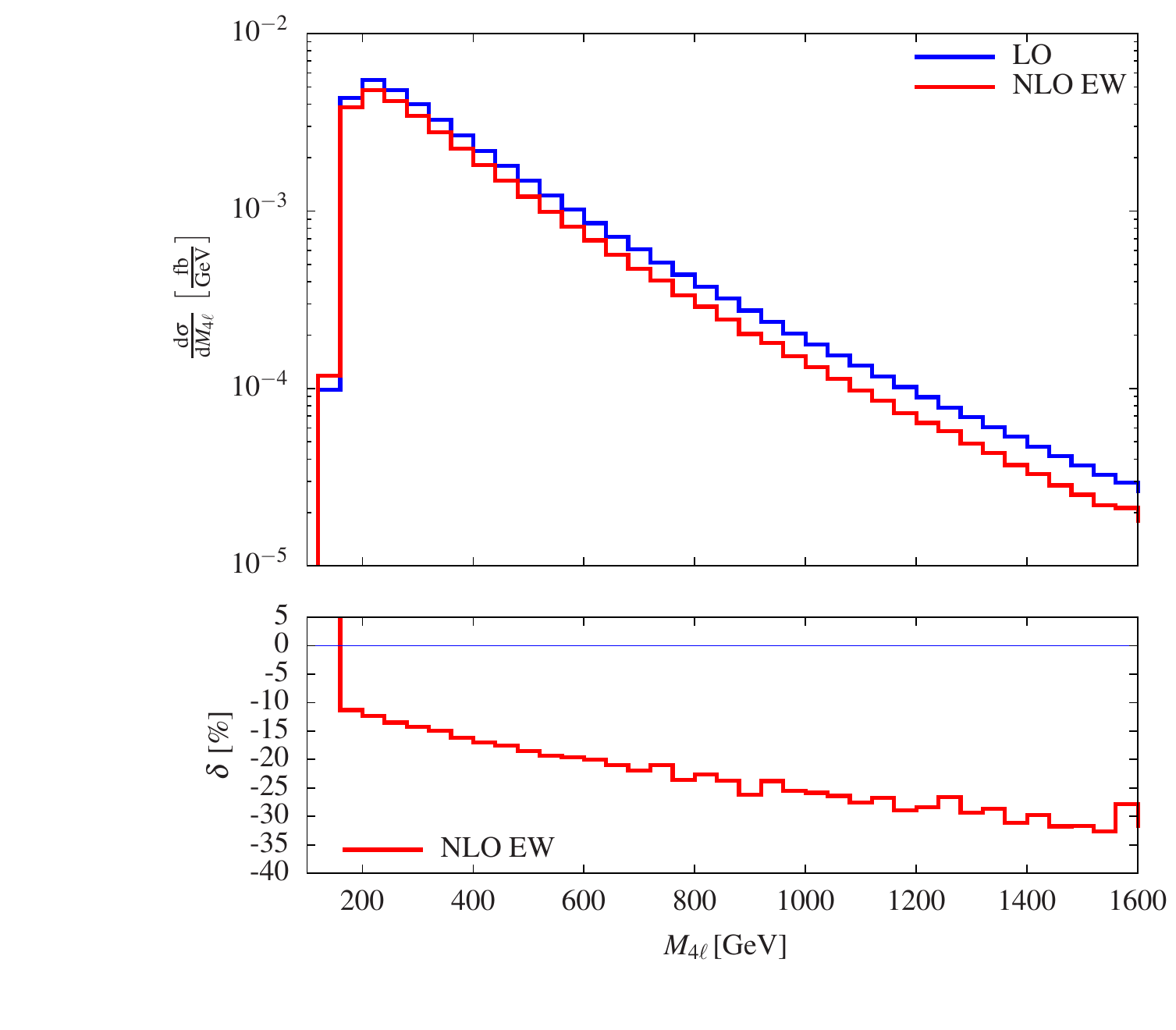}
\vspace*{-1em}
\caption{Invariant-mass distribution of the four leptons in $\Pp\Pp\to\mu^+\nu_\mu\Pe^+\nu_{\Pe}\Pj\Pj$
including NLO EW corrections (upper panel) and relative NLO EW corrections (lower panel).}
\label{fig:4l}
\end{figure}

In \reffi{fig:4l}, the distribution in the invariant mass of the four
leptons is displayed.  The upper panel shows the LO and NLO EW
prediction and the lower panel the relative EW corrections $\delta =
\sigma_{\NLO \; \EW} / \sigma_{\LO} - 1$ in per cent.  The cross
section drops by one order of magnitude only $500\GeV$ above its
maximum. For typical gauge-boson pair production processes like
$\PW\PW$ or $\PZ\PZ$ production the cross section decreases more than
twice as fast with increasing energy.  The negative EW corrections
increase from $-12\%$ at $150\GeV$ to $-32\%$ at $1.6\TeV$.

\begin{figure}
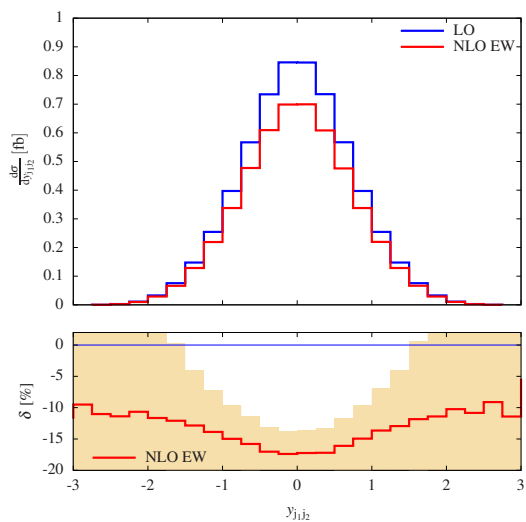

\hspace{-2cm}
\includegraphics[width=.47\textwidth]{{{histogram_rapidity_j1j2}}}
\vspace*{-1em}
\caption{Rapidity distribution of the leading jet pair in $\Pp\Pp\to\mu^+\nu_\mu\Pe^+\nu_{\Pe}\Pj\Pj$
including NLO EW corrections (upper panel) and relative NLO EW corrections (lower panel).
The yellow band describes the expected statistical experimental
uncertainty for a high-luminosity LHC collecting $3000\fb^{-1}$ and
represents a relative variation of $\pm 1/\sqrt{N_{\rm obs}}$ where $N_{\rm obs}$ is the number of observed events in each bin.}
\label{fig:rapj1j2}
\end{figure}

In \reffi{fig:rapj1j2}, the rapidity distribution of the dijet system
is presented.  In VBS the two jets are typically back to back, and
their joint rapidity tends to be close to zero.  Near $y_{\Pj_1\Pj_2}=
0$, the EW corrections are maximal and at the level of $-16\%$ as for
the integrated cross section.  
For large $|y_{\Pj_1\Pj_2}|$ the two jets tend to be in the same
hemisphere, and the kinematics is different from the one of VBS.
In this kinematic region, the EW corrections turn out to be smaller.
The variation of the EW corrections is weaker in
other rapidity distributions of the jets and practically absent in
those of the leptons.

In addition to the relative corrections, in the lower panel also the
expected statistical experimental uncertainty is displayed.  Here we
assume $3000\fb^{-1}$ which is the target for a high-luminosity LHC.
For each bin we have computed the number of expected events $N_{\rm
  obs}$ and the corresponding relative uncertainty as $\pm
1/\sqrt{N_{\rm obs}}$ in per cent which is represented by the band.
This clearly demonstrates that the EW corrections are mandatory to
describe VBS with sufficient precision at a high luminosity LHC.  Note
that the expected statistical experimental uncertainty for the total
cross section is $1.6\%$.

We follow the experimental analysis and do not include real radiation
of W and Z~bosons. Including these contributions with realistic
experimental cuts would only partially compensate the virtual
corrections \cite{Baur:2006sn} and even a fully inclusive treatment of
massive gauge-boson radiation would not yield a complete cancellation of
the Sudakov logarithms \cite{Ciafaloni:2001vt}.

\subsection{Origin of large electroweak corrections}

Upon splitting the EW corrections into the gauge-invariant subsets of
fermionic and bosonic parts, we could attribute the large effects
exclusively to the bosonic sector.  We have furthermore verified at
the level of distributions that the leading behavior of the NLO EW
corrections is dominated by the virtual corrections.  In order to get a
feeling for the relevant scales in the process, we calculated the
average partonic center-of-mass energy, the average invariant mass of
the jet pair, and the average invariant mass of the four-lepton
system and found $\langle\sqrt{\hat{s}}\rangle \sim 2.2\TeV$, $\langle
m_{\Pj\Pj} \rangle \sim 1.6\TeV$, and $\langle m_{4 \ell} \rangle \sim
390\GeV$, respectively.

In view of the complexity of the VBS process and the appearance of
many different scales, the study of approximations is useful in order
to understand the origin of the large corrections.

\begin{figure}
\includegraphics[width=.35\textwidth]{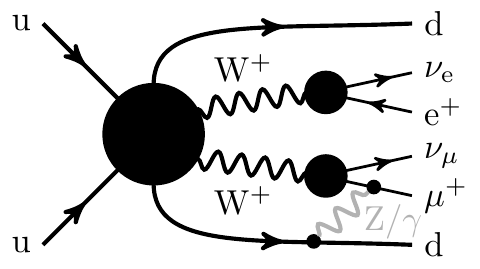}
\vspace*{-1em}
\caption{Schematic representation of the double-pole approximation for the VBS process.
The black blobs represent the factorized subprocesses while the grey gauge boson represents a non-factorizable correction.}
\label{fig:nloDPA}
\end{figure}

As a first step, we have evaluated the subtracted virtual corrections
to the VBS process in the double-pole approximation (DPA)
\cite{Dittmaier:2015bfe,Denner:2016jyo}. In the DPA two on-shell
W~bosons are requested, and the matrix elements are split into those
for production and decay of the resonant W~bosons.  This is
illustrated in \reffi{fig:nloDPA} where the blobs represent the
production and decay processes of the W~bosons including the
factorizable corrections, while the explicit neutral gauge boson
($\PZ,\gamma$) constitutes a typical non-factorizable correction in
this framework.  We found that the DPA reproduces the full process
within $1\%$ and thus provides a sufficiently good approximation for
practical purposes. In the DPA, the factorizable corrections
constitute $\sim 95\%$ of the subtracted virtual corrections and are
thus responsible for the large EW corrections. Moreover, the
non-factorizable corrections result exclusively from photon exchange
and are compensated upon adding the corresponding real photonic corrections.

\begin{figure}
\includegraphics[width=.35\textwidth]{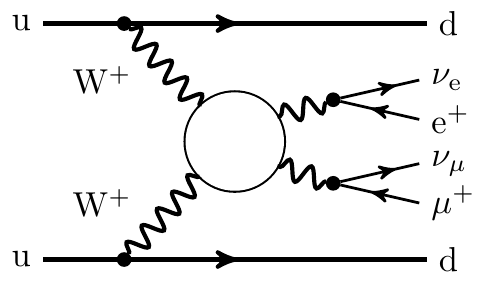}
\vspace*{-1em}
\caption{Schematic representation of the EVBA.
The white blob represents the VBS subprocess.}
\label{fig:EFTVBS}
\end{figure}

To further simplify the discussion, we use the EVBA, depicted schematically in \reffi{fig:EFTVBS}, where two W~bosons
are radiated off the quark lines to scatter.
In this picture, most of the energy is transferred to the two
back-to-back jets while the rest of the energy goes into the
scattering of the two W bosons in the central region.
The invariant masses of the radiated W~bosons are space-like but of
the order of the W-boson mass to enhance the cross section \cite{Altarelli:1987ue,Kuss:1995yv,Rauch:2016pai}. 
While the EVBA constitutes a crude approximation valid only in the very-high-energy limit
\cite{Kuss:1995yv,Accomando:2006hq}, it is sufficient to discuss the
origin of the enhanced EW corrections. 

To proceed, we combine the EVBA with the Sudakov approximation in a
similar way as pioneered in \citere{Accomando:2006hq} for VBS in
electron--positron annihilation. In the Sudakov limit, where all
invariants are large, the dominant EW corrections result from double
and single logarithms involving ratios of the large invariants and the
vector-boson masses squared \cite{Ciafaloni:1998xg,Denner:2000jv}. In
the EVBA the large logarithms in the factorizable corrections result
only from the vector-boson-scattering subprocess,
$\PW^+\PW^+\to\PW^+\PW^+$. To keep things simple, we only consider the
double EW logarithms, the collinear single EW logarithms, and the
single logarithms resulting from parameter renormalization.  Following
\citere{Denner:2000jv}, we obtain
\begin{align}
 \sigma_{\rm LL} = \sigma_{\rm LO} \biggl[1 & -\frac{\alpha}{4\pi} 4
   C^{\rm ew}_{\PW} \log^2\left(\frac{Q^2}{\MW^2}\right) \notag\\
& +\frac{\alpha}{4\pi} 2 b^{\rm ew}_{\PW} \log\left(\frac{Q^2}{\MW^2}\right) \biggr],
\end{align}
where the EW Casimir operator and the $\beta$-function coefficient for W~bosons read
\begin{eqnarray}
 C^{\rm ew}_{\PW} = \frac{2}{s_{\rm w}^2},\qquad
 b^{\rm ew}_{\PW} = \frac{19}{6s_{\rm w}^2}
\end{eqnarray}
with the sine of the weak mixing angle ${s_{\rm w}}$.  Using $\langle
m_{4 \ell} \rangle\sim 390\GeV$ as a typical scale $Q$ for the VBS
subprocess leads to an EW correction of about $-16\%$. Applying this
logarithmic approximation differentially to the distribution in the
invariant mass of the four leptons yields about
$-15\%$.  These numbers reproduce remarkably well the full correction of $-16\%$
given the fact that they include only logarithmic corrections
resulting from the VBS subprocess, neglecting even the
angular-dependent leading logarithms.

The resulting EW corrections are by a factor of 3--4 larger compared
to processes like vector-boson pair production or top-quark pair
production for the following reasons. First, the EW Casimir operator
$C^{\rm ew}$ is larger for vector bosons than for fermions. This
enhances the double logarithmic corrections by a factor of $1.5$ for
$\PW\PW\to\PW\PW$ with respect to $\Pq\bar\Pq\to\PW\PW$.  Second, the
typical scale of the hard scattering process $Q$ is larger for
$\PW\PW\to\PW\PW$.  For a typical pair-production process the scale is
more or less of the order of the pair production threshold, \ie\ 
$Q\sim250\GeV$ for $\Pq\bar\Pq\to\PW\PW/\PZ\PZ$, since the cross
sections drop with $1/\hat{s}$ above threshold. For
$\PW\PW\to\PW\PW$, on the other hand, the cross section drops much
slower (\emph{c.f.}~\reffi{fig:4l}) owing to the massive $t$-channel
vector-boson exchange in these processes \cite{Denner:1997kq}. Without
cuts, the cross section would even approach a constant for high
energies.  The scale $Q=\langle m_{4 \ell} \rangle\sim390\GeV$
enhances the double logarithmic corrections by a factor of $1.9$ with
respect to a scale $Q\sim250\GeV$. Third, the cancellation between
single and double logarithms is weaker for external vector bosons than
for external fermions.

\subsection{Summary}

In summary, the NLO EW corrections to a VBS process have been
presented for the first time.  The  fully differential computation includes
the full EW NLO matrix elements with all non-resonant and off-shell
contributions and can be well reproduced by a double-pole approximation.
The EW corrections to the fiducial cross section are with $-16\%$
surprisingly large and are an intrinsic feature of VBS.  Using the
effective vector-boson approximation combined with a high-energy logarithmic
approximation for the EW corrections, we have been able to identify
the source of these large corrections.  They result from large
logarithms in the virtual EW corrections to the VBS subprocess. They
are enhanced with respect to other LHC processes owing to the
comparably large couplings of the vector bosons, reduced cancellations
within the logarithmic corrections, and the massive
$t$-channel vector-boson exchange contributions in VBS, which lead to
sizeable contributions at large scales.
The large EW corrections presented here should be included in any precise analysis of VBS.

\vspace{1em}

\subsection{Acknowledgements}

We thank Jean-Nicolas Lang for providing a version of {\sc Recola}
featuring only fermionic corrections and Mauro Chiesa for useful discussions.
We acknowledge support by the German Federal Ministry for Education and Research (BMBF) under
contract no. 05H15WWCA1 and the German Science Foundation (DFG) under reference number DE~623/6-1.

\bibliography{vbs_ew}

\end{document}